\documentclass[sigconf, 10pt, nonacm]{acmart} 

\usepackage{graphicx}
\usepackage{url}
\usepackage{xcolor}

\title{The Cambridge Report on Database Research}

\usepackage{xstring}
\usepackage{etoolbox}

\begin{document}

\author{Anastasia Ailamaki}
\affiliation{%
 \institution{École Polytechnique Fédérale de Lausanne (EPFL)}
 \country{Switzerland}
}
\email{anastasia.ailamaki@epfl.ch}

\author{Samuel Madden}
\affiliation{%
 \institution{Massachusetts Institute of Technology}
 \country{USA}
}
\email{madden@csail.mit.edu}

\author{Daniel Abadi}
\affiliation{%
 \institution{University of Maryland}
 \country{USA}
}
\email{abadi@cs.umd.edu}

\author{Gustavo Alonso}
\affiliation{%
 \institution{ETH Zurich}
 \country{Switzerland}
}
\email{alonso@inf.ethz.ch}

\author{Sihem Amer-Yahia}
\affiliation{%
 \institution{CNRS, Université Grenoble Alpes}
 \country{France}
}
\email{sihem.amer-yahia@univ-grenoble-alpes.fr}

\author{Magdalena Balazinska}
\affiliation{%
 \institution{University of Washington}
 \country{USA}
}
\email{magda@cs.washington.edu}

\author{Philip A. Bernstein}
\affiliation{%
 \institution{Microsoft Research}
 \country{USA}
}
\email{philbe@microsoft.com}

\author{Peter Boncz}
\affiliation{%
 \institution{Centrum Wiskunde \& Informatica (CWI)}
 \country{Netherlands}
}
\email{boncz@cwi.nl}

\author{Michael Cafarella}
\affiliation{%
 \institution{Massachusetts Institute of Technology}
 \country{USA}
}
\email{michjc@csail.mit.edu}

\author{Surajit Chaudhuri}
\affiliation{%
 \institution{Microsoft Research}
 \country{USA}
}
\email{surajitc@microsoft.com}

\author{Susan Davidson}
\affiliation{%
 \institution{University of Pennsylvania}
 \country{USA}
}
\email{susan@cis.upenn.edu}

\author{David DeWitt}
\affiliation{%
 \institution{Massachusetts Institute of Technology}
 \country{USA}
}
\email{david.dewitt@outlook.com}

\author{Yanlei Diao}
\affiliation{%
 \institution{École Polytechnique and Amazon Web Services}
 \country{France}
}
\email{yanlei.diao@polytechnique.edu}

\author{Xin Luna Dong}
\affiliation{%
 \institution{Meta Reality Labs}
 \country{USA}
}
\email{lunadong@fb.com}

\author{Michael Franklin}
\affiliation{%
 \institution{University of Chicago}
 \country{USA}
}
\email{mjfranklin@uchicago.edu}

\author{Juliana Freire}
\affiliation{%
 \institution{New York University}
 \country{USA}
}
\email{juliana.freire@nyu.edu}

\author{Johannes Gehrke}
\affiliation{%
 \institution{Microsoft}
 \country{USA}
}
\email{johannes@microsoft.com}

\author{Alon Halevy}
\affiliation{%
 \institution{Google Cloud}
 \country{USA}
}
\email{alon.halevy@gmail.com}

\author{Joseph M. Hellerstein}
\affiliation{%
 \institution{UC Berkeley}
 \country{USA}
}
\email{hellerstein@berkeley.edu}

\author{Mark D. Hill}
\affiliation{%
 \institution{University of Wisconsin-Madison}
 \country{USA}
}
\email{markhill@cs.wisc.edu}

\author{Stratos Idreos}
\affiliation{%
 \institution{Harvard University}
 \country{USA}
}
\email{stratos@seas.harvard.edu}

\author{Yannis Ioannidis}
\affiliation{%
 \institution{University of Athens and Athena Research Center}
 \country{Greece}
}
\email{yannis@di.uoa.gr}

\author{Christoph Koch}
\affiliation{%
 \institution{École Polytechnique Fédérale de Lausanne (EPFL)}
 \country{Switzerland}
}
\email{christoph.koch@epfl.ch}

\author{Donald Kossmann}
\affiliation{%
 \institution{Microsoft Research}
 \country{USA}
}
\email{donald.kossmann@microsoft.com}

\author{Tim Kraska}
\affiliation{%
 \institution{Massachusetts Institute of Technology}
 \country{USA}
}
\email{kraska@csail.mit.edu}

\author{Arun Kumar}
\affiliation{%
 \institution{University of California, San Diego}
 \country{USA}
}
\email{akk018@ucsd.edu}

\author{Guoliang Li}
\affiliation{%
 \institution{Tsinghua University}
 \country{China}
}
\email{liguoliang@tsinghua.edu.cn}

\author{Volker Markl}
\affiliation{%
 \institution{Technische Universität Berlin}
 \country{Germany}
}
\email{volker.markl@tu-berlin.de}

\author{Renée Miller}
\affiliation{%
 \institution{University of Waterloo}
 \country{Canada}
}
\email{rjmiller@uwaterloo.ca}

\author{C. Mohan}
\affiliation{%
\country{USA} 
}
\email{seemohan@gmail.com}

\author{Thomas Neumann}
\affiliation{%
 \institution{Technical University of Munich}
 \country{Germany}
}
\email{neumann@in.tum.de}

\author{Beng Chin Ooi}
\affiliation{%
 \institution{National University of Singapore}
 \country{Singapore}
}
\email{ooibc@comp.nus.edu.sg}

\author{Fatma Ozcan}
\affiliation{%
 \institution{Google, Inc}
 \country{USA}
}
\email{fozcan@google.com}

\author{Aditya Parameswaran}
\affiliation{%
 \institution{UC Berkeley}
 \country{USA}
}
\email{adityagp@berkeley.edu}

\author{Ippokratis Pandis}
\affiliation{%
 \institution{Amazon Web Services}
 \country{USA}
}
\email{ippo@amazon.com}

\author{Jignesh M. Patel}
\affiliation{%
 \institution{Carnegie Mellon University}
 \country{USA}
}
\email{jigneshp@andrew.cmu.edu}

\author{Andrew Pavlo}
\affiliation{%
 \institution{Carnegie Mellon University}
 \country{USA}
}
\email{pavlo@cs.cmu.edu}

\author{Danica Porobic}
\affiliation{%
 \institution{Oracle Corporation}
 \country{Switzerland}
}
\email{danica.porobic@oracle.com}

\author{Viktor Sanca}
\affiliation{%
 \institution{Oracle Corporation}
 \country{USA}
}
\email{viktor.sanca@oracle.com}

\author{Michael Stonebraker}
\affiliation{%
 \institution{Massachusetts Institute of Technology}
 \country{USA}
}
\email{stonebraker@csail.mit.edu}

\author{Julia Stoyanovich}
\affiliation{%
 \institution{New York University}
 \country{USA}
}
\email{stoyanovich@nyu.edu}

\author{Dan Suciu}
\affiliation{%
 \institution{University of Washington}
 \country{USA}
}
\email{suciu@cs.washington.edu}

\author{Wang-Chiew Tan}
\affiliation{%
 \institution{Facebook AI}
 \country{USA}
}
\email{wangchiew@meta.com}

\author{Shiv Venkataraman}
\affiliation{%
 \institution{Google}
 \country{USA}
}
\email{shiv@google.com}

\author{Matei Zaharia}
\affiliation{%
 \institution{UC Berkeley}
 \country{USA}
}
\email{matei@berkeley.edu}

\author{Stanley B. Zdonik}
\affiliation{%
 \institution{Brown University}
 \country{USA}
}
\email{sbz@cs.brown.edu}

\newcommand{\myCommaSeparatedAuthors}{%
  Anastasia Ailamaki, Samuel Madden, Daniel Abadi, Gustavo Alonso, Sihem Amer-Yahia, Magdalena Balazinska, Philip A. Bernstein, Peter Boncz, Michael Cafarella, Surajit Chaudhuri, Susan Davidson, David DeWitt, Yanlei Diao, Xin Luna Dong, Michael Franklin, Juliana Freire, Johannes Gehrke, Alon Halevy, Joseph M. Hellerstein, Mark D. Hill, Stratos Idreos, Yannis Ioannidis, Christoph Koch, Donald Kossmann, Tim Kraska, Arun Kumar, Guoliang Li, Volker Markl, Renée Miller, C. Mohan, Thomas Neumann, Beng Chin Ooi, Fatma Ozcan, Aditya Parameswaran, Ippokratis Pandis, Jignesh M. Patel, Andrew Pavlo, Danica Porobic, Viktor Sanca, Michael Stonebraker, Julia Stoyanovich, Dan Suciu, Wang-Chiew Tan, Shiv Venkataraman, Matei Zaharia, and Stanley B. Zdonik%
}

\twocolumn[{
  \begin{center}
    {\LARGE\bfseries The Cambridge Report on Database Research \par}
    \vskip 1em
    {\large \myCommaSeparatedAuthors\par}
    \vskip 2em
  \end{center}
}]


\renewcommand{\shortauthors}{Ailamaki et al.}


\section{Introduction}

On October 19–20, 2023, the authors of this report convened in Cambridge, MA, to discuss the state of the database research field\footnote{Broadly defined as the community of researchers publishing in ACM SIGMOD, VLDB, and related conferences, journals, and workshops.}, its recent accomplishments and ongoing challenges, and future directions for research and community engagement. This gathering continues a long-standing tradition in the database community, dating back to the late 1980s, in which researchers meet roughly every five years to produce a forward-looking report~\cite{10.1145/1060710.1060718, 10.1145/1462571.1462573, 10.1145/2845915, 10.1145/306101.306137, 10.1145/3385658.3385668, 10.1145/382272.1367994, 10.1145/242223.242295, 10.1145/122058.122059, 10.1145/381854.381886}.

This report summarizes the key takeaways from our discussions. We begin with a retrospective on the community’s academic, open-source, and commercial successes over the past five years. We then turn to future opportunities, with a focus on core data systems—particularly in the context of cloud computing and emerging hardware—as well as on the growing impact of data science, data governance, and generative AI.

This document is not intended as an exhaustive survey of all technical challenges or industry innovations in the field. Rather, it reflects the perspectives of senior community members on the most pressing challenges and promising opportunities ahead.

\section{Evolution Over The Past Five Years}

The past five years have continued to see important advances in the database and data systems landscape, particularly around new hardware, cloud-based data systems, and the continued adoption of statistical techniques, ML, and AI in both core data systems architecture and components. 

The rise of Large Language Models (LLMs) has significantly shaped the collective consciousness of both computer science and society in recent years. While LLM-related technologies are still evolving and have yet to reach their full potential, they offer a promising solution to many complex data challenges, particularly those involving natural language and unstructured data. Already, LLMs have unlocked new possibilities for understanding human intentions and needs, paving the way for more intuitive, natural language-based querying and analysis interfaces. They have also demonstrated the capacity to comprehend data, including unstructured formats such as video and text, and to ground structured data in broader general knowledge. Additionally, LLMs are capable of synthesizing complex, multi-step data transformation programs. If fully realized, these technologies promise to revolutionize the ability of data systems to understand users, data, and programs. This has already prompted researchers to reconsider traditional database interactions, broadening the scope to incorporate unstructured data and natural language into conventional database systems. We explore these LLM-related opportunities in greater detail in Section~\ref{sec:llm} below.

\subsection{Research Successes}

In this section, we briefly review some of the key areas of progress in the community in the past few years.

\subsubsection{Core Data Systems}
In reaction to the low-level MapReduce-style tools of the previous decade of Big Data, database research and products make great strides toward usability and rich functionality of databases at massive scales. In particular, cloud-native architectures have matured significantly, and the industry has widely adopted the concept of disaggregated storage and compute, enabling a high degree of scalability and flexibility.

The hardware landscape continues to evolve rapidly to cater to resource-hungry AI, opening up new challenges and opportunities for data systems. The database community has made strides in leveraging improved hardware capabilities, such as NVMe SSDs, GPUs, DPUs, and specialized AI accelerators such as FPGAs and ASICs. For example, research on NVMe SSDs has led to the development of new storage engines that can fully utilize their high IOPS and low latency, often redesigning traditional data structures such as B-trees to minimize random accesses. Work on persistent memory has resulted in novel index structures that provide crash consistency without the overhead of traditional write-ahead logging. In the realm of specialized accelerators, researchers have explored using FPGAs for operations such as data decompression and filtering and GPUs for massively parallel query processing, particularly for tasks such as hash joins and sorting. Looking forward, the memory hierarchy continues to evolve, with new generations of SSDs and changes in interconnect technologies impacting database system design. The advent of CXL (Compute Express Link) has sparked research into memory expansion and sharing techniques across servers. These hardware advancements also raise intriguing new questions about optimal data placement and query execution models in this increasingly heterogeneous hardware environment.

\subsubsection{ Human Centric Systems and Data Science}
Data sharing and collaboration have become increasingly important, driven by the need to break down data silos and enable cross-organizational analytics. The popularity of ``data lake'' platforms that host data from many different organizational silos presents opportunities and challenges in terms of privacy, governance, and query processing across distributed datasets. This also demands novel methods to enable data discovery at the dataset level that support high-level analytics tasks including deriving data explanations and improving machine learning models.

More broadly, related fields of data science and data engineering have also continued to grow, with an increasing focus on end-to-end data pipeline and workflow systems that seamlessly  data preparation, analysis and visualization, and ML/AI. These pipelines are used both in an exploratory mode, where they are iteratively developed and refined, as well as for the deployment of live services.
These systems have given rise to increasingly sophisticated tools for workflow and data pipeline management, data discovery, integration and cleaning, synthetic data generation, metadata and log management, and code and data versioning.

\subsubsection{ ML and AI for Data Systems}
Machine learning for DBMS internals has progressed from a promising idea to practical implementations with several important applications. For example:

\begin{itemize}
\item In query optimization, cost models based on learning over data and workloads demonstrate superior performance when compared to traditional cost-based optimizers, especially for complex, multi-join queries; 

\item Cardinality estimation advances with increasingly sophisticated deep neural net and statistical models that can capture high-dimensional correlations in data distributions; 

\item Reinforcement learning techniques are deployed to dynamically adjust physical data organization based on observed query workloads, with predictive I/O techniques demonstrating the potential to outperform traditional indexing methods;

\item In related fields, such as cloud resource management, ML models for serverless VM management have shown significant reductions in cold start times and resource costs. 
\end{itemize}

Although ML/AI-based methods do not always outperform other approaches, the community is learning how to scope these research ideas for fruitful adoption in practical settings.

\subsubsection{Responsible Data Management}
Finally, the rise in the prevalence of AI models for interpreting data and making complex decisions has led to an awareness of the need for research on responsible data management, effectively integrating data management research into the area of responsible AI (RAI). Responsible data management is motivated by the observation that the decisions we make during data collection and preparation profoundly impact the accuracy, fairness, robustness, interpretability, and legal compliance of AI systems. In other words, responsible data management is foundational to RAI. Over the last decade, responsible data management has been integrated into the mainstream research agenda in data management, as evidenced by numerous keynotes, tutorials, and research papers appearing in leading conferences and journals.
This development mirrors the integration of RAI into research agendas throughout computer science and data science, in communities including core AI, ML, and NLP, as well as the establishment of highly visible dedicated venues including ACM FAccT, ACM/AAAI AIES, and ACM Transactions on Responsible Computing.

\subsection{Commercial and Open Source Successes}

The database field continues to have a strong impact on the industry, with several notable commercial successes in recent years. Cloud-based DBMS services have seen explosive growth, with major cloud providers now offering various solutions, from traditional relational DBMSs to NoSQL and distributed SQL (previously NewSQL) offerings. A significant trend has been toward systems built on shared storage, where any processing node can access any data element with a soft allocation of nodes to data partitions; this is in contrast to the ``shared-nothing'' approach that had been the dominant architecture for distributed database designs in previous decades, where data is physically partitioned between processors. The shared-access design facilitates independent scaling of compute and storage and allows for a separation of concerns between the data processing and durable storage layers. Services such as Amazon Redshift, Google BigQuery, Databricks, Microsoft Fabric, and Snowflake employ this approach. Even transactional DBMSs have adopted this model, with systems such as Azure SQL and Amazon Aurora moving to disaggregated architectures.

Open-source database engines with roots in research continue to see increased adoption and commercialization. Almost four decades after the initial Postgres research system, PostgreSQL continues to be a popular choice for both on-premises and cloud deployments, offering a robust, feature-rich, and extensible database system that continues to validate the long-term impact of academic open-source database technology. Apache Spark and Apache Flink are further examples of scalable data processing systems that originated in academia and have become widely adopted open-source systems. DuckDB, an embeddable analytical database engine, has gained attention for efficient query processing in local environments and is increasingly popular in contexts where database engines were previously absent, including data science workflows and graphical user interfaces. Its architecture incorporates advanced techniques including vectorized execution and compressed columnar storage, demonstrating how database research can be translated into practical, widely accessible tools.

Another development in database system architecture has been the emergence of composable building blocks in projects such as Velox, DataFusion, and Calcite. These approaches aim to create reusable, high-performance components for data processing that can be easily integrated into different systems. By standardizing interfaces for common database operations and components and providing optimized implementations, the development of new data systems for a variety of use cases is enabled through a flexible, reusable, and modular architecture of individually optimizable and replaceable components without the need to redesign the entire system.

Finally, the rise of data science has spurred the development of new systems bridging traditional data management and the needs of data scientists. DBMS-style semantics and optimization have been brought to bear on data science workloads. Systems such as Dask, Modin, Polars, and Spark showcase how principles from distributed DBMSs and query optimization can be applied to new kinds of workloads, expanding the reach and impact of database research.

\subsection{Community Status and Health}

Expanding our community into artificial intelligence and machine learning has been a defining trend. Many database researchers are now working at the intersection of data management and AI, exploring ML for query optimization, AI-powered data integration, efficient data systems for AI workloads, user interfaces for human-AI collaboration on tasks including data wrangling, querying and metrics, and data-centric tools and benchmarks to improve the AI lifecycle. This expansion reflects the growing importance of AI in data-intensive applications and the perspective that database researchers can bring to these challenges. It has also led to increased collaboration with AI/ML communities, fostering cross-pollination of ideas and techniques. Flagship DB conferences such as VLDB and SIGMOD have also launched new research tracks/categories focused on scalability and data management issues in data science/AI/ML, which have been well received. Given the central role of data management in data-driven exploration, it is natural for our research to straddle different areas and impact multiple application areas and scientific domains. As a step to recognize work that applies data management research to address practical problems, SIGMOD and VLDB now invite application papers that describe such efforts. 

Several initiatives aimed at promoting diversity and inclusion in the community have been launched. Notably, DBCares provides a mechanism for addressing harassment issues and promoting a more inclusive environment at our conferences. We have also seen efforts to improve the diversity of conference attendees, speakers, and organizers. The DEI in DB Initiative (\url{https://dbdni.github.io/}), a first-of-its-kind effort among computing research communities, has streamlined such efforts across multiple DB conferences, enabling sharing of knowledge, best practices, and measurement of these efforts' impact over time. Finally, the addition of video captioning for conference talks has improved access to our presentations. 
We return to some of these community challenges in Section~\ref{sec:community} below.

\section{New Challenges and Looking Forward}

The coming years present a number of research challenges for the data systems community, and there continue to be many opportunities to innovate in the design of data systems, which we address in this section. Challenges, questions, and opportunities arise in a number of key areas, including:
\begin{itemize}
\item Managing the complexity that arises from the ongoing migration of enterprises to the cloud and the proliferation of cloud data systems, 
\item The proliferation of new, heterogeneous hardware, with computation moving closer to storage and networks and ever more sophisticated accelerators for AI and other workloads,
\item The sudden rise of generative AI and LLMs, and the associated challenges, which potentially change how we interface with relational databases and allow us to build data management systems for much less structured data,
\item Data governance and the responsible management of data and our role in these topics,
\item The interaction between data management and ``data science'', and how our community can contribute to the ecosystem of data science tools.
\end{itemize}

We describe each of these areas in more detail below.

\subsection{Core Data Systems}

Our meeting broadly covered four areas of core data systems:  classical database system architectures, the movement to the cloud, opportunities in new hardware, and applications of (non-LLM/generative AI) AI/ML techniques to data systems.

\subsubsection{Database Systems}

A key research direction involves building extensible databases or databases out of reusable components.  One important opportunity we discussed is how to architect databases to support this type of extensibility -- for example, should we aim for POSIX-style specs for the interfaces between components (i.e., optimizer, storage manager, etc)? If successful, such efforts could allow more ``mixing and matching'' of components from different vendors and systems. Some efforts to modularize data systems already exist (e.g., Velox, DataFusion, Calcite).  More provocatively, the rise of agent-based designs where different components of complex systems interact through natural language (NL) could be applied to data systems; in such a setting, for example, a query optimizer might receive an NL description of a query, and translate it into an NL operator plan which an executor agent consumes. As with interfaces to systems, it is unclear if NL is useful for its expressivity and generality or merely a source of ambiguity when specifying system interfaces. 
Other opportunities related to interfaces that we discussed include the deeper integration of applications and database systems, e.g., by developing modern front ends that go beyond JDBC/ODBC and include proxy functionalities as well as application hints.

A second opportunity is federation. As data platforms become more federated, with data spread across multiple engines and locations, new architectural paradigms will emerge. These systems will be capable of pulling data from diverse sources to answer complex queries, posing new challenges around automated infrastructure management and performance optimization. For instance, techniques to push down query predicates to remote data sources can help to enhance performance by reducing data movement. Addressing issues related to write propagation and data integration in such environments will be essential to ensure consistency and reliability.

Federation is increasingly important, especially in regulated environments where data residency laws require data not to leave specific jurisdictions, such as the EU.  Designing data systems to support this distributed form of model execution will be increasingly important and involves tackling challenges such as data heterogeneity and cross-border data transmission while ensuring data integrity and security protocols meet regulatory standards.

Finally, with the increasing diversity of compute and storage hardware, many opportunities abound to improve the design of data systems to take advantage of this hardware. We discuss these opportunities along with some of the hardware trends below.

\subsubsection{Cloud Data Systems}

Cloud computing has fundamentally changed both the architecture and the commercial landscape of databases and database systems in the last decade, with many enterprises having moved to the cloud and the majority of database system revenue now coming from cloud computing. This shift has brought new systems and commercial players, including Amazon, Azure, Google Cloud, Databricks, Snowflake, and many more, as well as a movement towards a more disaggregated storage architecture, where data resides in a highly reliable cloud file system and data systems build on top of this layer. Data systems built for this environment need a different set of optimizations -- to manage, for example, the lower bandwidth and higher latency of cloud storage vs. local SSDs -- as well as a different set of designs, for example, to leverage a high degree of reliability and replication provided by cloud storage.

We identified several opportunities and challenges in cloud computing, including automated infrastructure management, taking advantage of new hardware inside data centers, and working with cloud vendors to develop benchmarks, representative traces, and better measures of resource consumption.

One interesting opportunity that we discussed focused on automated infrastructure management, where components of cloud computing systems are themselves automatically managed. A theme was control plane management, i.e., self-healing, auto-provisioning, and always-up features, where many interesting advancements have been made and will continue to evolve.

Some research directions here include:

\begin{itemize}
\item Database virtualization, where a single database front-end automatically provisions and routes queries to the best infrastructure,

\item Declarative infrastructure as a service, where declarative interfaces are used for specifying more than queries but also the infrastructure upon which systems run, with search and optimization systems that attempt to allocate infrastructure in the most cost-effective way.
\end{itemize}

Another opportunity is collecting high-quality data to benchmark cloud systems and train machine learning models that will power adaptive and learned features of systems. Existing benchmarks including TPC-DS or workload traces like the IMDB benchmark don't capture the diversity and complexity of real-world workloads or the variable load of cloud systems. Some cloud companies are starting to address this by releasing anonymized or aggregated workloads from their operational environments, which involves both capturing query logs and instrumenting systems to gather telemetry information about usage patterns, performance metrics, and anomalies. Such data collection itself is complicated as it requires anonymizing queries and logs. An important area of future research will be creating synthetic benchmarks that accurately mirror real-world use cases. Here, LLMs may offer a way forward as they can perhaps be asked to generate synthetic but similar queries; it is, however, unclear if this can be done in a way that does not leak private information.

A related topic is how academic researchers can best contribute to cloud-based data systems, given that almost all cloud-based databases are controlled by a few closed-source commercial vendors. A widely used open-source cloud-native database system would offer an important proving ground for research experimentation as well as an alternative to commercial vendor lock-in.

\subsubsection{New Hardware and Green Computing}

The data management community has long recognized the importance of optimizing data systems for evolving hardware to enable broader, more efficient use. As the pace of hardware scaling slows and demand for data-driven applications grows, there is a pressing need for deeper collaboration across the software-hardware stack—including with architects, OS developers, and language designers. The shift toward specialized compute, storage, and networking is particularly relevant, with the rise of accelerators such as GPUs, TPUs, FPGAs, and SmartNICs offering opportunities to optimize data workflows. Key research directions include developing abstractions to leverage diverse accelerators, exploiting parallelism in GPUs for database operations, and even designing new data-centric accelerators. Placement and connectivity of these accelerators—via technologies similar to CXL and UCIe—will also shape how data systems are architected for performance and flexibility.

At the same time, memory remains a critical bottleneck, both in cost and scalability, especially as processor core counts grow and DRAM scaling plateaus. Emerging solutions such as CXL-enabled memory tiers, pooled or shared memory configurations, and processing near memory (e.g., in SmartNICs or DPUs) offer promising directions but require significant software adaptation. Utilizing older GPUs for general-purpose data tasks and better managing GPU memory and data transfer costs are additional areas of exploration. 

Finally, the environmental impact of data systems -- including energy usage and the embodied carbon cost of hardware, e.g., data storage and RAM -- remains an under-addressed issue within the community. As data center power demands continue to grow, incorporating sustainability as a first-class design principle is an important challenge for the future.

\subsubsection{Learned Components, Autotuning, and Opportunities for ML-in-databases}

There is a continuing opportunity related to integrating learning and data systems, focused on how ML can help achieve cost efficiency and simplicity in modern data platforms. For instance, over the past several years ML models have been shown to be able to optimize data layouts dynamically, determining the most efficient way to store data based on access patterns. Similar models can be applied in many areas—for example, to choose the best I/O strategies based on learned estimates of selectivities. These capabilities extend into other areas such as admission control and resource allocation, with the promise of offering a fine-grained, adaptive approach to resource utilization, whether in a serverless or provisioned environment. Determining the best models for each of these areas and integrating them into database systems in a way that provides real-world efficiency gains and is robust to both data and workload shifts are key challenges.


More generally, AI/ML techniques will continue to see adoption in database system components that need to adapt to specific data or workload distributions -- particularly components that traditionally relied upon heuristics and/or coarse-grained statistics. This includes slowly-changing phenomena like learning data distributions for cardinality estimation, where it pays off to periodically train complex models. It also includes quickly-changing phenomena e.g., scheduling or resource allocation where it makes sense to use lightweight adaptive mechanisms like reinforcement learning. AI/ML techniques have seen less uptake in practical settings where conventional algorithms and data structures work well; in these settings, the potential benefits from data-specific learned strategies are often outweighed by challenges in tuning and debugging.

In the cloud, auto-tuning will also play an increasingly important role for maintaining and optimizing databases throughout their lifecycle. These include ML models capable of deciding when to provision additional resources, when to scale resources up or workloads down, when to upgrade system versions, and when to migrate workloads or when to shift load to specialized engines that are more scalable or have special acceleration for certain types of workloads. In addition, with the rapid deployment of GPUs and other new hardware, there will increasingly be opportunities for cloud systems to ``burst out'' load to this new hardware when appropriate. 

Relatedly, data systems will need to be modified to facilitate control through ML models; existing parameter tuning may be sufficient but new parameters may be needed, e.g., to support shipping of load to other systems, partial execution of queries on specialized hardware, etc. Properly devising parameters can offer substantial manageability gains.

\subsection{Generative AI / Large Language Models}
\label{sec:llm}

GenAI and especially LLMs have quickly become the hottest topic in computer science. These technologies are likely to have a lasting impact outside of obvious applications in automated code generation and conversational agents / chatbots -- especially as they pertain to large scale database and data processing applications. 

 Clearly, LLMs will play a key role in the ease of management by enabling more intuitive human interfaces for complex database systems. These models can help users interact with data in a natural language, lowering the barrier for entry and making data systems more user-friendly. Furthermore, LLMs can assist in auto-generating queries, optimizing them for performance, and even suggesting schema designs based on workload patterns and business requirements. However, it's unclear whether LLMs are the right solution for many classic DBMS problems – given their current computation cost, it seems unlikely, for example, that it will make sense to use LLMs to solve query optimization problems, or that APIs between data systems components will ever be replaced by ``agents'' interacting via natural language.

More broadly, from a data perspective, new AI models offer an opportunity to extend the scope of data processing to images, text, video, large documents collections, and more. It is no longer science fiction to imagine ``querying'' a corpus of thousands of documents or images to identify entities, trends, numbers, and figures. We can apply declarative database interface and principles to these new domains. 

Advancing LLM integration presents several key opportunities, including building efficient computational stacks by combining traditional techniques -- e.g.,  data partitioning, caching, and embedding indexes -- to handle the high cost of inference at scale. Reducing hallucinations remains a top priority, with databases and provenance tools playing a crucial role in validating outputs. Decomposing complex tasks into simpler sub-tasks can boost accuracy, aligning well with the community’s prior work on declarative queries and crowd-sourcing. There is also a need to move beyond basic RAG methods by developing smarter, context-aware retrieval systems. Finally, ensuring LLMs correctly interpret user questions calls for better interfaces and human-in-the-loop validation.

We discuss these opportunities in data systems in more detail, focusing on how LLMs might be integrated into data processing engines and how our community can contribute data systems for enabling the AI/LLM ecosystem.

\subsubsection{LLMs for Database Systems}

LLMs already excel at some database and data systems-specific tasks and problems. For example, they are better than previous models at the text-to-SQL problem, and their variants are consistently at the top of task leaderboards, such as the Spider Text-To-SQL Challenge (\url{https://yale-lily.github.io/spider}). However, even in this case, where LLMs present a seeming breakthrough on non-public datasets, such as traditional data warehouses, their performance has been much less impressive unless special techniques can be applied, such as using prior workloads or human/expert knowledge in the prompt.

Research on LLMs for database systems spans multiple open questions and directions. One key challenge is how LLMs can assist in database tasks beyond the well-studied text-to-SQL interface. They may help reduce human effort by interpreting database manuals, tuning database parameters, and aiding DBAs. 
As database systems grow more complex and LLM inference techniques improve, their use in online tasks such as query execution may become feasible. However, a single LLM inference is unlikely to suffice for complex tasks, necessitating robust pipelines that integrate verification steps and human oversight to ensure accuracy and reliability.


LLMs must also learn to interact with database APIs, adapting to different system interfaces through prompting or in-context learning. Furthermore, their effectiveness in handling relational and other structured data remains an open question. Currently, even state-of-the-art models struggle with fundamental relational properties, such as the set-based nature of relations. Developing models specifically trained for set-oriented data could improve efficiency, but maintaining such models must be balanced against the rapid evolution of general-purpose LLMs. Additionally, cross-modal embeddings may enhance LLMs' ability to process relational, textual, and novel data types e.g.,  time series and nested tables containing heterogeneous types.


Another challenge involves the future of SQL and alternative query languages. Conversational interfaces and query debuggers will be crucial to helping users validate and trust LLM-generated queries. Given that LLMs can hallucinate or make mistakes, effective explanation mechanisms, human-LLM collaboration, and schema-independent validation methods will become increasingly important.

Finally, beyond query processing, LLMs could play a role in broader system design tasks, such as composing database engines and designing data pipelines. Allowing LLMs to understand large codebases and contribute to system development would be an intriguing area for further exploration.

\subsubsection{Data Systems for LLMs}

The database community can also contribute significantly to building data systems that support LLMs. As LLM architectures continue to expand, it will be increasingly important to develop efficient infrastructures to manage large multi-modal data sets, optimize fine-tuning, and ensure scalability, fault tolerance, and elasticity in native cloud environments of both training and serving systems. Future systems may need to accommodate new storage and access methods as LLMs span diverse modalities, including text, code, images, video, and audio.

Another critical area is data quality, labeling, and metadata management for LLM training, as well as creating evaluation frameworks for LLM applications ("evals" for short). The quality of pretraining data differentiates newer LLM generations, yet large-scale datasets contain both valuable content and low-quality or biased data. Effective tools are needed to filter, de-bias, and curate such data efficiently. The database community’s experience in data exploration and summarization can be adapted to address these challenges, particularly in the context of multi-modal datasets. Likewise, reliable programmatic evals remain a roadblock for LLM applications to graduate from prototypes to production. Effective tools are needed to reliably post-process output data, debug failures, and adapt inputs to LLM application workflows by leveraging both human feedback and emerging reinforcement learning methods at scale.

Optimizing inference and prompt engineering workflows also presents an opportunity. Efficient inference algorithms, hardware-aware techniques, and standardized benchmarks will be essential to improving performance. In addition, multi-step reasoning tasks often require complex prompt engineering workflows involving metadata management and strategic data chunking. While current practices are largely ad hoc, new tools such as LangChain and LlamaIndex suggest that structured database management techniques could enhance workflow organization, governance, and auditability.

 Retrieval-augmented (RAG) systems offer another set of research challenges. As vector databases and feature stores become integral to LLM applications, questions arise about their fundamental differences from traditional RDBMSs. Approximate nearest neighbor (ANN) search has been developed to handle billion-scale embedding datasets, but better techniques are needed to handle very high-dimensional vectors and combine ANN search with other filter conditions. 
 Furthermore, as LLMs incorporate more multi-modal data, novel indexing and retrieval methods may be necessary.

Finally, beyond one-shot inference and retrieval-augmented systems, complex ``agentic'' AI workflows are emerging that integrate multiple LLM inferences, retrieval steps, ML models, and external tools such as code executors or search engines. These compound AI systems improve functionality, robustness, and efficiency but introduce new trade-offs in latency and accuracy. Developing frameworks that help optimize these workflows -- balancing factors including correctness, consistency, cost, and performance -- are a key opportunity. Drawing inspiration from traditional database transaction models, new abstractions and query optimization techniques may be needed to manage these AI-driven pipelines effectively.

\subsection{Responsible Data Management and Data Governance}

Incorporating ethics and legal compliance into data-driven systems requires a broader, lifecycle-oriented perspective that extends beyond the “last mile” of fair and interpretable machine learning. A truly responsible data or AI system must consider both the data and system life cycles—from data provenance and validity to design goals, deployment impacts, and unintended consequences. This calls for novel algorithmic approaches centered on responsibility objectives and a reimagined data engineering and model development stack that supports tasks for responsible data management and continuous oversight. The database community, with its deep expertise in data lifecycle management, is perhaps a natural choice to lead this transformation.

This is increasingly pressing as the rise of large-scale AI models and centralized data repositories has intensified the need for robust data governance, including privacy, security, and regulatory compliance. Our community can contribute tools including metadata management and data catalogs that are essential for enabling data discoverability, sharing, and monetization across organizations. As data continues to grow in value, especially for training LLMs, our community must develop new methods for federated data management, privacy-preserving sharing, and interoperable standards. These efforts also need to consider issues around responsible data management that encompass fairness, robustness, interpretability,  throughout the  AI lifecycle.

Key research gaps in responsible data management include developing methods to assess data quality in relation to specific downstream tasks and socially meaningful metrics e.g., fairness, accuracy, and robustness; creating techniques to improve data quality through acquisition, cleaning, and preprocessing; establishing lifecycle-aware provenance tracking to meet the diverse interpretability needs of stakeholders such as data scientists, auditors, and affected individuals; and designing educational resources to help researchers and engineers consider the context of use and stakeholder perspectives in building and operating data-intensive systems.

 




\subsection{Collaboration, Integration, and Human-Centric Data Issues}

There are a number of data systems problems around how to allow users to find, integrate, and interact with data. Some of these problems -- such as data integration -- have been long-standing challenges for our community, while others -- such as our community's role in platforms for data sharing and data science -- have become more pressing in recent years.

\subsubsection{Data Sharing and Collaboration}

Data sharing and the rise of data set search, data markets, data lakes, and other tools and platforms for exchanging data have become increasingly common. Indeed, data sharing has been a theme of prior database community meetings. There are many aspects to these problems, including systems issues including scale and distribution, as well as issues for privacy, security, veracity, copy detection, and data ownership. The rise of large generative AI models has raised some new issues and opportunities, particularly as organizations amass massive training data sets used to power these models. 

We identified several promising research questions in this space, including:
\begin{itemize}
\item With the rise in data volumes, how do we enable easy and accurate discovery of relevant data and discard or avoid useless information? Particularly with LLMs and RAG architectures, when too much data is indexed, much of what is retrieved is useless or irrelevant.


\item What are the tradeoffs around data accumulation and retention as organizations accumulate more data? There are potential negatives around privacy and the environment, but new models demand more data. With ever-increasing data volumes, how do we ensure that we are only collecting and retaining what is needed?

\item How do we level the playing field around data accessibility? Large organizations are accumulating immense volumes of data, giving them an advantage when making predictions or training models. In many cases, there is also a tension between accessibility and privacy and questions about who has a right to retain data. Similar issues arise with regard to operational data and benchmarks of data systems when a few large vendors control the systems and workloads.
\end{itemize}

\subsubsection{Data Integration}

Data integration remains a significant challenge, even with the advent of Large Language Models (LLMs). While LLMs have shown promise in various areas of data management, their ability to fully solve the complex problem of data integration is still uncertain and requires further investigation. The potential for automated interoperability through catalogs and LLMs is an area of active research. These technologies may also offer new approaches to SQL dialect translation.

Simply having a data lake does not equate to having integrated data. While data lakes provide a centralized repository for diverse data types, true integration requires addressing semantic differences, resolving entity-matching issues, and ensuring data quality and consistency across sources. These challenges persist even in the context of modern data architectures.

The database community continues to grapple with fundamental data integration issues such as schema matching, entity resolution, and data cleaning at scale. While LLMs and other AI technologies may help to address these challenges, they also introduce new complexities, such as the need to understand and mitigate potential biases in AI-assisted integration processes. Moreover, there is still the challenge of scale: How would an LLM be able to tell if two tables are related in a data lake with thousands of tables, each with millions to billions of tuples? In most cases, it is impossible to tell based on schema alone, requiring inspection of the data itself. Simply feeding every record in a data lake into an LLM prompt is infeasible, so ways to expose just enough data to facilitate integration are needed. As we explore the potential of LLMs and other emerging technologies in data integration, we must also continue to advance our understanding of the semantic and structural aspects of data integration. This includes developing more sophisticated methods for inter-operating between natural language and structured data,  capturing and representing data context, improving techniques for automated schema mapping, and creating more robust frameworks for assessing and ensuring data quality in integrated environments.

\subsubsection{Human-Centered Systems}

As AI and LLMs become increasingly integrated into data systems, the database community’s longstanding focus on human-centered approaches is more vital than ever. We believe our goal should be to build systems that augment human ability to manage and analyze data while addressing the limitations of LLMs—such as hallucinations—through mechanisms including fact-checking, database constraint maintenance, and user-verified results. Revisiting work on interactive GUIs and incorporating ideas from crowd-sourcing, such as task decomposition and consensus strategies, can help balance AI-generated outputs with human judgment and oversight. Supporting users in familiar tools, e.g., spreadsheets, BI platforms, and computational notebooks remains critical, with opportunities to enhance these environments using database concepts such as indexing, declarative queries, and automated optimization.

There is also a potential to leverage LLMs for tasks such as automated insight discovery and visualization, though scalability and relevance to end-users remain challenges. Designing intuitive, natural-language-driven interfaces and explanatory tools will be key to making complex data tasks more accessible to non-experts. Emphasizing human-in-the-loop feedback and continuous learning mechanisms will help systems evolve with user needs and build trust. Collaboration with HCI, visualization, and programming languages communities—who bring valuable expertise in usability and program synthesis—will be essential. By grounding AI-driven data systems in human-centered design, the community can develop trustworthy, intelligent tools that empower users and ensure the effectiveness and accessibility of future data management solutions.

\subsubsection{Data Science and Data-Intensive Science}

The database community continues to work to bring database-inspired optimizations—such as query planning and efficient execution—to popular data science, AI/ML, and visualization tools. Projects like Modin aim to scale dataframe libraries like pandas using familiar database techniques. However,  the idea of a single, universal language or paradigm (e.g., extending SQL) covering all data programming needs is unlikely, due to the diversity and specialization of data science tasks. Instead, efforts should focus on developing interoperable systems that allow different tools and languages to work together more efficiently, enhancing performance while respecting domain-specific workflows. This includes identifying core data models and operations, handling deviations like nested or non-atomic types, and leveraging LLMs to assist with multi-language data pipeline construction.

At the same time, there is a tension between improving existing widely used tools and advocating for cleaner or higher performance abstractions that may have a steeper adoption curve. While making current tools like pandas more scalable remains important, there’s also potential in defining more streamlined, learnable, and optimized dataframe abstractions that unify ideas from tools like Dask, Polars, Ibis, and Spark. In parallel, domain-specific languages tailored to particular data science workflows—such as ML data loading or cleaning—are being developed with embedded database principles. Overall, we should strive to enhance usability, scalability, and performance throughout the data analytics pipeline, not by forcing a one-size-fits-all solution, but by improving integration and optimization across the ecosystem of data tools.

\section{Database Community Status and Health}
\label{sec:community}

In this section, we briefly review several issues that were raised regarding the community's status in terms of community and conference structure, research velocity, inclusivity, and community impact.

\subsection{Community Structure}

One challenge facing our growing community is how to maintain focus and relevance in light of more papers and intensive growth of interest in AI and ML. For example, in the last 4 years, SIGMOD submissions grew by 96\% -- nearly doubling. The SIGMOD program of accepted papers grew by 66\%, with 100 more papers in '25 than in '22.





We have an influx of borderline papers on AI-related topics, many of which appear better suited to AI/ML conferences (and some were likely rejected from those conferences). Introducing some low-effort back-pressure is necessary, to prevent a flood of judgment-call papers that reviewers and chairs must assess for scope.
Another challenge is  to recruit and manage hundreds of reviewers per conference per year, and do a responsible job reviewing. Issues include that (a) it is difficult to assemble a PC made up of people whose reputations the chairs knew firsthand, (b) there are inevitably some reviewers who do subpar work, and (c) the burden placed on reviewers and chairs by multiple submission rounds and the sheer volume of submissions often leads to burnout by the end of the process.

As our conferences grow, the conference experience has become less effective in bringing the community together. SIGMOD, VLDB, ICDE, EDBT, and other conferences  have many parallel activities (often five or more!), which results in a community siloed into narrow specializations. While there has been experimentation with fewer tracks and not having all papers presented at SIGMOD each year, parts of the community strongly oppose any ``quality tiers'' that differentiate accepted papers like the AI community has adopted. This raises the question: What are the goals of conferences and physical collocation, and how do we scale our conferences to achieve those goals?

In addition, there are more general issues that face computer science broadly, including collusion and inclusivity. With respect to collusion, there are serious concerns about groups of authors forming ``collusion rings'' where reviewers aim to get each other's papers accepted in CS publishing/reviewing broadly
If the community wants to address this problem, it needs well-supported software to allow for network analysis of connectivity (co-authorship, co-citation, co-location) between authors and reviewers. Right now this is done in  ad hoc ways.

Finally, while our community has improved inclusivity through efforts including DBCares, the DEI in DB Initiative, video captioning, and more, there remain questions about how we can broaden participation and make it an inviting place for all. Some specific suggestions that were discussed include standardizing on double-blind reviewing to reduce the risk and perception that more senior authors' papers may be treated favorably and ensuring that under-represented groups  are not asked to do excessive service or reviewing work in an effort to be inclusive.

\subsection{Education and Relevance in an AI-Dominated Landscape}

The rapid rise of AI and LLMs presents challenges and opportunities for database education. We must adapt curricula to teach students how to use AI tools such as LLMs effectively, focusing on analytical thinking and problem-solving rather than rote syntax. Although LLMs have transformed some aspects of database education, many core data management and efficient data systems engineering concepts remain foundational. Our focus should shift towards developing higher-order thinking skills that LLMs cannot replicate.

The excitement around AI is leading to a decrease in graduate student interest in traditional CS fields, including databases and data systems. To address this, we must highlight the ongoing importance of these systems in enabling AI applications and emphasize the exciting research challenges at the intersection of databases and AI. Moreover, expanding our pedagogy to include data engineering and broader ``data-oriented thinking'' will better prepare students for the evolving data landscape.

\subsection{Incentives for Impactful Research}

The database community has a strong tradition of collaboration with industry, ensuring our research remains practical and impactful. However, as AI applications gain prominence, we must ensure that database research maintains its relevance and impact.

The growing global community and increasing volume of publications has led to a broader and more diverse research landscape. However, this also means that many young researchers  struggle to navigate the landscape and identify impactful work. To address this, senior community leaders must set clear guidelines and encourage a focus on high-impact, real-world research applications.

Finally, we discussed the importance of developing incentives for impactful research that go beyond paper submissions to include contributions, including valuable software, datasets, and reproducible results. Concrete suggestions include new award schemes that recognize impactful work across research approaches, including algorithmic, theoretical, and systems-based contributions and encouraging open research, including negative results, that further incentivize impactful, long-term contributions and contribute to the collective knowledge of the community.

\section{Conclusion}
We hope this report will serve to guide the database community's research priorities in the coming years. We remain a lively, strong, and opinionated group of dedicated and close-knit  academic and industrial researchers and engineers. Our core mission remains building data management systems, addressing new and emerging challenges in doing so for the cloud, for emerging hardware, and for and with generative AI. We are optimistic that we have many lasting contributions to look forward to in the years to come.

\section*{Acknowledgments}
We are grateful to Google for their financial support of the Cambridge Database Meeting. We also would like to acknowledge the MIT student scribes who took notes at the meeting: Tianyu Li, Matthew Russo, Sivaprasad Sudhir, Geoffrey Yu, Sylvia Zhang, and Xinjing Zhou.

\bibliographystyle{ACM-Reference-Format}
\bibliography{references}

\end{document}